\documentclass[aps,pra,twocolumn,groupedaddress,showpacs,10pt]{revtex4-1}

\usepackage{graphicx}
\usepackage{dcolumn}
\usepackage{bm}
\usepackage{setspace}
\usepackage{color}
\usepackage{overpic}
\usepackage{sidecap}
\usepackage{braket}
\usepackage{amsmath}
\usepackage[english]{babel}
\begin{document}

\title{Optical visibility and core structure of vortex filaments in a bosonic superfluid}

\author{Franco Dalfovo$^1$}
\author{Russell N.~Bisset$^1$}
\author{Carmelo Mordini$^{1,2}$}
\author{Giacomo Lamporesi$^{1,2}$}
\author{Gabriele Ferrari$^{1,2}$}

\affiliation{$^1$INO--CNR BEC Center and Dipartimento di Fisica, Universit\`a di Trento, 38123 Trento, Italy}
\affiliation{$^2$Trento Institute for Fundamental Physics and Applications, INFN, 38123, Trento, Italy}

\begin{abstract}

We use optical images of a superfluid consisting of  a weakly interacting Bose-Einstein condensate of sodium atoms to investigate the structure of quantized three-dimensional vortex filaments. We find that the measured optical contrast and the width of the vortex core quantitatively agree with the predictions of the Gross-Pitaevskii equation.

\end{abstract}

\maketitle

\section{Introduction}

The Gross-Pitaevskii (GP) equation was independently derived by L.P.~Pitaevskii \cite{Pitaevskii61} and E.P.~Gross \cite{Gross61} in 1961. It describes a superfluid gas of weakly interacting bosons at zero temperature. The solution of the equation is a complex function $\Psi=|\Psi| \exp i \varphi$, whose  modulus squared represents the particle density, $n=|\Psi|^2$, and the gradient of the phase gives the local velocity of the fluid, ${\bf v}= (\hbar/m)\nabla \varphi$, where $m$ is the particle mass.  In the derivation by L.P. Pitaevskii, the GP equation emerges as a generalization of Bogoliubov's theory \cite{Bogolyubov47} to a spatially inhomogeneous superfluid \cite{Ginzburg58}.  A quantized vortex can exist as a stationary solution of the GP equation where all particles circulate with the same angular momentum $\hbar$ around a line where the density vanishes; the solution has the form $\sqrt{n(r)} \exp i \varphi$, where now $\varphi$ is the angle around the vortex axis and $r$ is the distance from the axis in cylindrical coordinates.  The density $n(r)$ is a smooth function which increases from $0$ to a constant asymptotic value $n_0$ over a length scale characterized by $\xi$, known as the {\it healing length}, determined by $n_0$ and the strength of the interaction. 

Quantized vortices have been extensively studied in superfluid $^4$He \cite{Donnelly91}, which is a strongly correlated liquid. The core of the vortex in $^4$He is only qualitatively captured by the GP equation and more refined theories are needed to account for the atom-atom interactions and many-body effects \cite{Dalfovo92,Ortiz95,Vitiello96,Giorgini96,Galli14}. A direct comparison between theory and experiment for the structure of the vortex core is not available, and is likely unrealistic, the main reason being that the core size in $^4$He is expected to be of the same order as the atom size. The only way to observe such a vortex thus consists of looking at its effects on the motion of impurities that may be attached to it. Electrons \cite{Yarmchuk79,Yarmchuk82,Maris09,Maris11}, solid hydrogen particles \cite{Bewley06,Bewley08,Paoletti10,Paoletti11,Fonda14}, and $^4$He$_2^*$ excimer molecules \cite{Zmeev13} have been used for this purpose. These impurities act as tracers for the position of vortex filaments in order to infer their motion on a macroscopic scale, but the fine structure of the core remains inaccessible. Furthermore, impurities may themselves affect the dynamics of the vortex filaments \cite{Barenghi09}. 

In dilute ultracold atomic gases the situation is more favorable. On the one hand, the GP theory furnishes a very accurate description of the system in regimes of temperature and diluteness that are attainable in typical experiments with trapped Bose-Einstein condensates (BECs) \cite{Dalfovo99,PSbook16}. On the other hand, beginning with a series of seminal experiments 
\cite{Matthews99,Madison00,Anderson2000,Anderson2001,Haljan01,AboShaeer01,Hodby02},
quantized vortices are routinely produced and observed with different techniques (see \cite{Fetter09} for a review). 

\begin{figure*}
\centering
\includegraphics[width=0.9\textwidth]{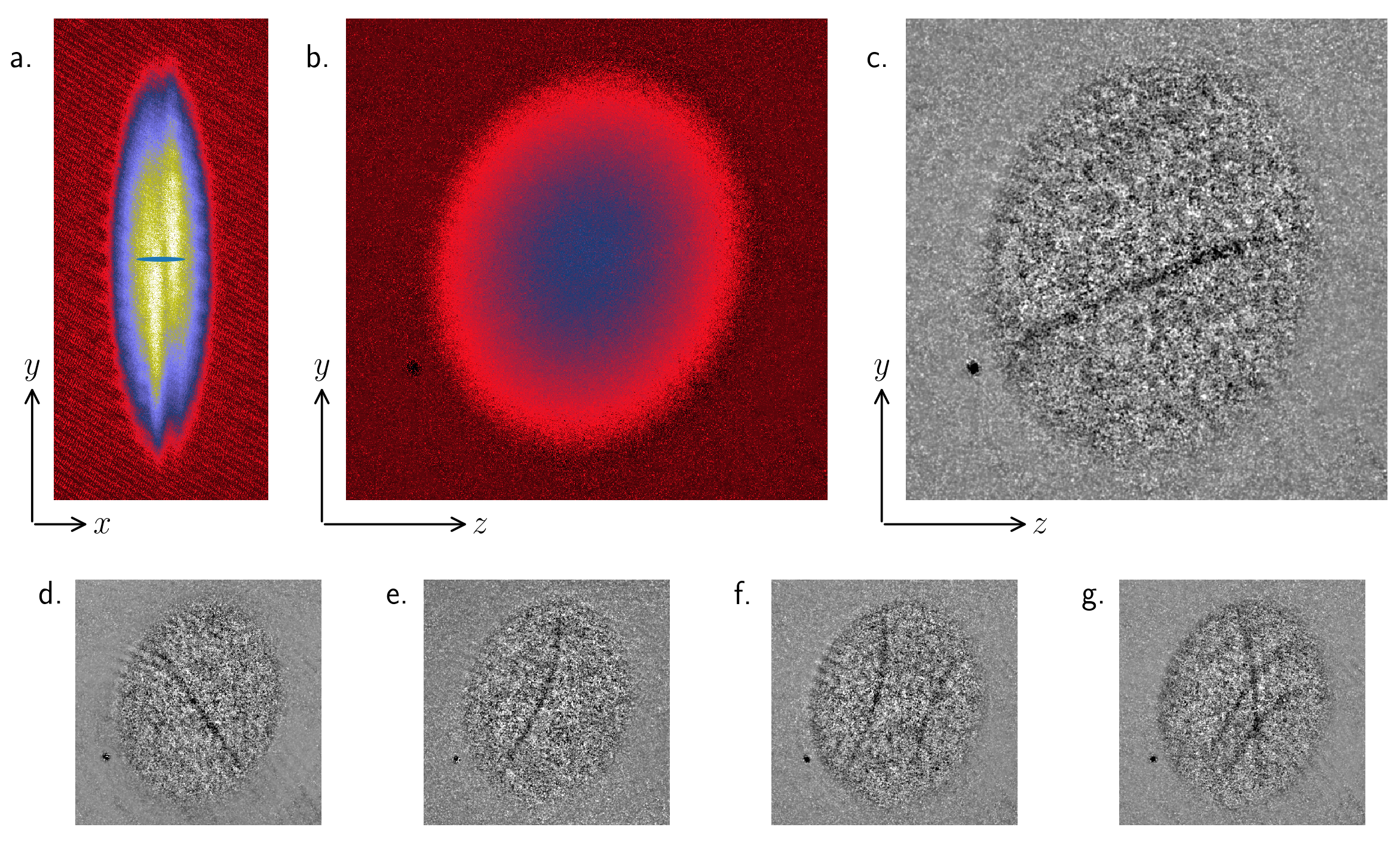}
\caption{Experimental absorption images of a condensate with $7 \times 10^6$ atoms after $120$~ms of free expansion. The small blue ellipse at the center of (a) represents the shape of the trapped condensate before the expansion, which is an elongated ellipsoid with the long axis in the $x$-direction. The expansion is faster in the transverse direction, so that the aspect ratio is inverted and the atomic distribution aquires a pancake shape. (a) Column density along a transverse direction. The faint vertical stripe is a signature of the presence of a vortex, and its shape is an interference pattern originating from the anisotropic velocity field around the vortex and the velocity field of the expansion. The  field of view is $1.3 \times 3$~mm. (b) Column density along the axial direction. The vortex is almost invisible. The field of view is $3 \times 3$~mm. (c) Residual column density. From the previous image we subtract spurious interference fringes, due to imperfections in the optical imaging, and the background density, using a Thomas-Fermi fit (see text). The result is an image of the residual column density which neatly reveals a vortex filament. (d)-(g) Other examples of vortex filaments shown by the residual column density for different condensates with one or more vortices. Note that even though the {\it in-situ} condensate is always isotropic in the $y$-$z$ plane it becomes slightly elliptic after a long expansion due to a residual curvature of the magnetic field used to levitate the condensate against gravity.}
\label{fig:expt-images}
\end{figure*}

Despite such an abundance of work, it may sound surprising that no detailed quantitative comparison between theory and experiment for the
structure of the vortex core in three-dimensional (3D) condensates has yet been performed. A reason is that the healing length $\xi$ in typical trapped BECs, though much larger than in liquid $^4$He, is still smaller than the optical resolution, which is limited by the wavelength of the laser beams used for imaging. Another reason is that, when illuminating the atomic cloud with light, the result is the optical density, which is determined by an integral of the density along the imaging axis (column density); thus, a vortex filament has a strong contrast only if it is rectilinear and aligned along the imaging axis. One can overcome the first limitation by switching off the confining potential, letting the condensate freely expand. The vortex core expands as well, at least as fast as the condensate radius \cite{Lundh98,Dalfovo00,Hosten03,Teles13a}, so that it can become visible after a reasonable expansion time.  Concerning vortex alignment, one can strongly confine a BEC along one spatial direction, squeezing it to within a width of several $\xi$. In such a geometry, vortices orient themselves along the short direction, thus behaving as point-like topological defects in a quasi-2D system rather than filaments in a 3D fluid (a recent discussion about the structure of the vortex core in expanding quasi-2D condensates can be found in \cite{Sang13}). Conversely, if the condensate width is significantly larger than $\xi$ in all directions, the vortex filaments can easily bend \cite{Aftalion01,Garcia01,Aftalion03,Modugno03}, with a consequent reduction of their visibility in the column density. Bent vortex filaments have indeed been observed in \cite{Raman01,Rosenbusch02,Bretin03,Donadello14}.
Bending and optical resolution particularly limit the quality of comparisons between theory and experiment for the structure of the vortex core (see Fig.~14.10 in \cite{PSbook16}). 

In this work, we show that 3D vortex filaments can be optically observed with enough accuracy to permit a direct comparison with the predictions of the GP theory. In our experiment, we produce large condensates of sodium atoms in an elongated axially symmetric harmonic trap and we image each condensate, in both the axial and a transverse direction, after free expansion. When a vortex filament is present, it produces a visible modification of the column density distribution of the atoms. 
We use numerical GP simulations, as well as scaling laws which are valid for the expansion of large condensates, to make direct comparisons with our experimental observations and find good agreement. 

\section{Experiment}

We produce ultracold samples of sodium atoms in the internal state $|3S_{1/2},F=1,m_{\mathrm{F}}=-1\rangle$ in a cigar-shaped harmonic magnetic trap with trap frequencies $\omega_x/2 \pi=9.3$~Hz and $\omega_\perp/2 \pi=93$~Hz.  The thermal gas is cooled via forced evaporative cooling and pure BECs of typically around $10^7$ atoms are finally obtained with negligible thermal component. The evaporation ramp in the vicinity of the BEC phase transition is performed at different rates: slow quenches eventually produce condensates which are almost in their ground state, while faster quenches lead to the formation of quantized vortices in the condensate as a result of the Kibble-Zurek mechanism \cite{Lamporesi13,Donadello16}.  The quench rate can be chosen in such a way to obtain condensates with one vortex on average. 

The trapped condensate has a radial width on the order of $30 \ \mu$m and an axial width that is $10$ times larger. The healing length in the center of the condensate is about $0.2\ \mu$m, smaller than the optical resolution. It is also about two orders of magnitude smaller than the radial width of the condensate, which means that, as far as the density distribution is concerned, a vortex is a thin filament living in a 3D superfluid background with smoothly varying density, and the local properties of the vortex core are hence almost unaffected by boundary conditions. However, boundaries are still important for the superfluid velocity field. In fact, the ellipsoidal shape of the condensate causes a preferential alignment of the vortex filament along a (randomly chosen) radial direction so as to minimize its energy. Moreover, this geometry makes the flow around the vortex line anisotropic, meaning that on the larger scale of the entire condensate a vortex behaves as an almost planar localized object. For this reason, such vortices in elongated condensates are also known as solitonic-vortices \cite{Donadello14,Brand02,Komineas03,Ku14}. For our purposes, such localization is an advantage since it significantly reduces the bending of the vortex filaments, while at the same time keeping their local core structure three dimensional. 

Observations are performed by releasing the atoms from the trap and taking simultaneous absorption images of the full atomic distribution along the radial and axial directions after a sufficiently long expansion in free space, so that the vortex core becomes larger than the imaging resolution \cite{Donadello14,Donadello16}. The presence of a levitating magnetic field gradient makes it possible to achieve long expansion times preventing the BEC  from falling. Typical images are shown in Fig.~\ref{fig:expt-images}.  In the radial direction (panel {\it a}), the vortex is seen as a dark stripe. This soliton-like character is due to the interference of the two halves (ends) of the elongated condensate which, on the large length scale of the entire condensate, have approximately a $\pi$ phase difference \cite{Donadello14,Ku14,Tylutki15}.
If a vortex filament is parallel to the imaging direction, the dark stripe exhibits a central dip, corresponding to the vortex core seen along its axis, and a twist due to the anisotropic quantized circulation. The $2\pi$ phase winding around the vortex core was also detected in the same setup \cite{Donadello14} by means of an interferometric technique based on a sequence of Bragg pulses. In the axial direction (panel {\it b}), the soliton-like character is integrated out and the vortex filament is only a faint (and almost invisible) perturbation in the column density. However, by subtracting the background represented by a condensate without any vortex, the filament clearly emerges in the residual density distribution (panel {\it c}). In the following we show how this signal can be used to extract quantitative information on the vortex structure after expansion, and how this is related to the shape of the vortex core in the condensate {\it in-situ}, before the expansion. 

\section{Theory}

The GP equation for the macroscopic wave function $\Psi({\bf r,t})$ for a BEC of weakly interacting bosons of mass $m$ at zero temperature is \cite{Pitaevskii61,Gross61,Dalfovo99,PSbook16}
\begin{equation}
i \hbar \frac{\partial \Psi}{\partial t} = \left( -\frac{\hbar^2 \nabla^2}{2m} + V_{\rm ext} + g |\Psi|^2 \right) \Psi ,
\end{equation}
where $V_{\rm ext}$ is the external potential and $t$ is time. The quantity $g$ is a coupling constant characterizing the interaction between the atoms, which is positive for our condensates.  The stationary version of the GP equation is obtained by choosing $\Psi({\bf r},t)= \psi({\rm r}) \exp (-i\mu t/\hbar)$, so that
\begin{equation}
\left( -\frac{\hbar^2 \nabla^2}{2m} + V_{\rm ext} + g |\psi|^2 \right) \psi = \mu \psi
\label{eq:stationaryGP}
\end{equation}
where $\mu$ is the chemical potential and  $n=|\psi|^2$ is the density. In our case, we use the stationary GP equation to describe the condensate confined by the axially symmetric harmonic potential $V_{\rm ext}=(m/2)[\omega_x^2 x^2+\omega_\perp^2 (y^2+z^2)]$, with the aspect ratio $\lambda=\omega_x/\omega_\perp=0.1$, as in the experiment.  Then we simulate the expansion by using this solution as the $t=0$, starting condition for the solution of the time dependent GP equation with $V_{\rm ext}=0$. We simulate condensates with and without a vortex. In the former case, the vortex is rectilinear, passing through the center and aligned along the $z$-axis. The need to accurately describe the dynamics of the system on both the scale of the healing length $\xi$ and the scale of the width of the entire expanding condensate poses severe computational constraints. With this in mind, we are only able to perform simulations up to values of the chemical potential on the order of $10\hbar \omega_\perp$, which are smaller than the experimental values, ranging from about $15$ to $30\hbar \omega_\perp$. Experiments can also be performed for smaller values of $N$, and hence smaller $\mu$, but fluctuations in the density distribution become relatively larger with decreasing $N$, and the signal-to-noise ratio for the visibility of vortices in axial imaging becomes too small. The comparison between theory and experiments hence requires an extrapolation of the GP results to larger $\mu$ and this is possible thanks to scaling laws which are valid for large condensates.  

If $\mu$ is significantly larger than both $\hbar \omega_\perp$ and  $\hbar \omega_x$, then the ground state of the condensate, i.e., the lowest energy stationary solution of the GP equation, is well approximated by the Thomas-Fermi (TF) approximation, which corresponds to neglecting the first term in the parenthesis of Eq.~(\ref{eq:stationaryGP}), so that the density becomes  \cite{Dalfovo99,PSbook16}
\begin{equation}
n_{\rm TF}(x,y,z) =  \frac{1}{g} \left[\mu -\frac{1}{2}m \omega_x^2 x^2 -\frac{1}{2}m \omega_\perp^2 (y^2+ z^2) \right] 
\label{eq:TF}
\end{equation}
within the central region where $n_{\rm TF}$ is positive, and is $0$ elsewhere.
We can then define the boundary TF radii $R_x = (2\mu/m\omega_x^2)^{1/2}$ and  $R_\perp = (2\mu/m\omega_\perp^2)^{1/2}$, the central density $n_0=\mu/g$, and the rescaled coordinates $\tilde{x}=x/R_x$, $\tilde{y}=y/R_\perp$ and $\tilde{z}=z/R_\perp$, and rewrite the density in the form 
\begin{equation}
n_{\rm TF}(\tilde{x},\tilde{y},\tilde{z}) =  n_0 ( 1 - \tilde{x}^2 - \tilde{y}^2 - \tilde{z}^2 ) \, .
\label{eq:rescaledTF}
\end{equation}
This inverted parabola is a very good approximation for the density profiles of our condensates except in a narrow region near the condensate boundaries  \cite{Dalfovo96b}. 

In the regime where the TF approximation is valid, the free expansion is governed by simple scaling laws \cite{Castin96,Kagan96,Dalfovo97}. In particular, one can prove that the condensate preserves its shape with a rescaling of the TF radii in time according to $R_x(t)=b_x(t)R_x(0)$ and $R_\perp(t)=b_\perp(t)R_\perp(0)$, where the scaling parameters $b_x$ and $b_\perp$ are solutions of the coupled differential equations
$\ddot{b}_\perp -  \omega_{\perp}^2/(b_xb_\perp^3)  =  0 $ and $\ddot{b}_x -  \omega_{x}^2/(b_x^2b_\perp^2)   = 0$,
with initial conditions $b_x=b_\perp=1$ and $\dot{b}_x=\dot{b}_\perp=0$ at $t=0$.
By using the aspect ratio $\lambda$ and introducing the dimensionless time $\tau=\omega_{\perp}t$, one can rewrite the same equations as 
\begin{equation}
\frac{d^2 b_\perp}{d \tau^2}  -  \frac{1}{b_xb_\perp^3} =  0 \ \ \ , \ \ \ 
\frac{d^2 b_x}{d \tau^2} -  \frac{\lambda^2}{b_x^2b_\perp^2}  =  0  \, . 
\label{eq:b}
\end{equation}
Analytic solutions exist in the limit $\lambda \ll 1$, that is, for a very elongated ellipsoid, for which one finds \cite{Castin96}
\begin{eqnarray}
b_\perp (\tau) & = &  \sqrt{1+\tau^2} \nonumber \\
b_x (\tau) & = & 1 + \lambda^2 [ \tau \ {\rm arctan} \tau - \ln \sqrt{1+\tau^2} \ ] \, .
\label{eq:banalytic} 
\end{eqnarray}
The correction proportional to $\lambda^2$ becomes vanishingly small in the limit of the infinite cylinder, where the condensate is known to follow a scaling behavior that preserves its radial shape, even in regimes where the TF approximation does not apply \cite{Pitaevskii97}.

\begin{figure}[]
\includegraphics[width=0.9\linewidth]{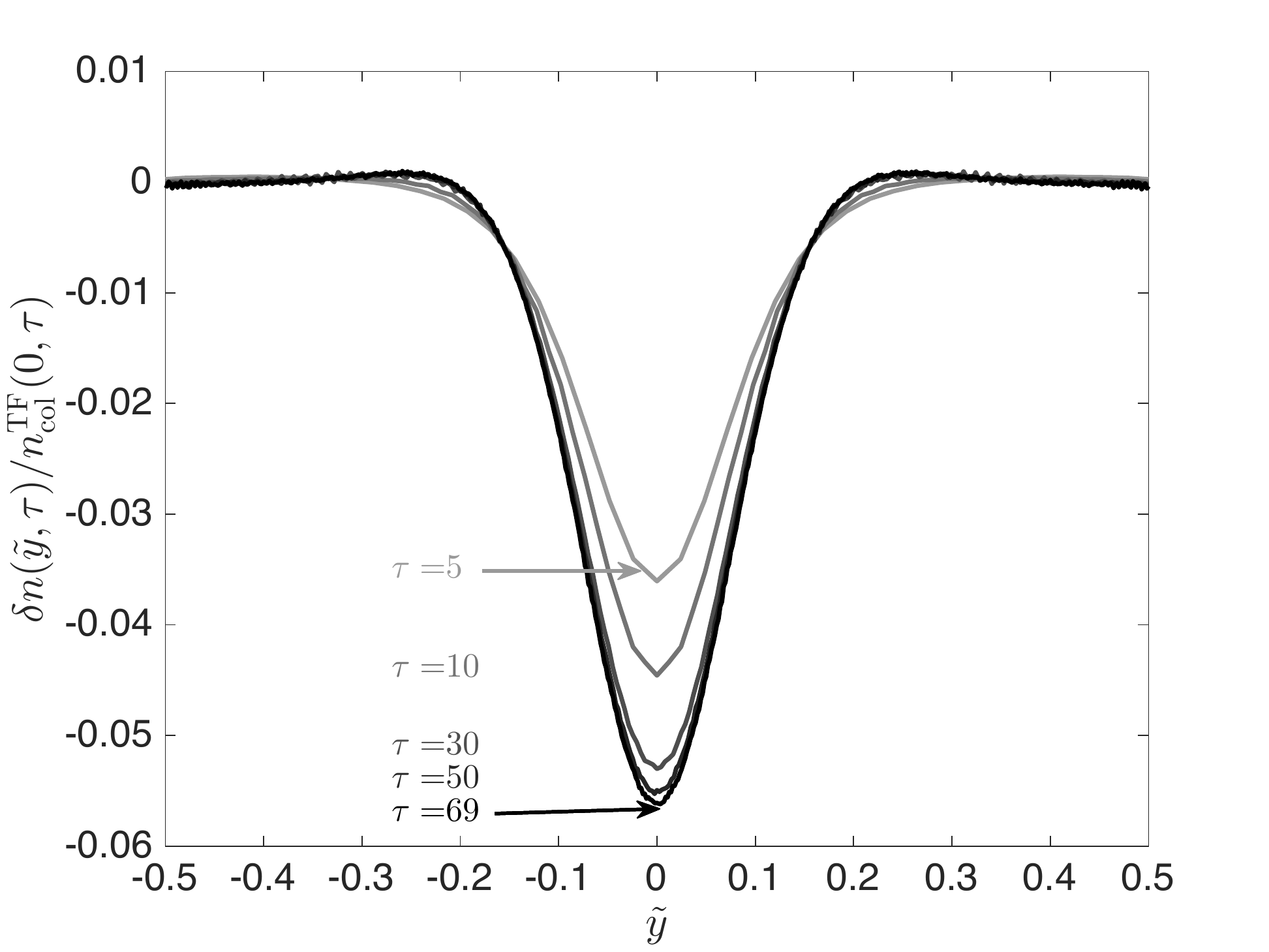}
\caption{Residual column density  (\ref{eq:nres}) calculated for a GP simulation of an expanding condensate with $\mu=9.7 \hbar \omega_\perp$ and with a vortex aligned along $z$, passing through the origin. Curves are plotted for different values of the expansion time, $\tau=\omega_\perp t$, and are normalized to the value $n^{\rm TF}_{\rm col} (0,\tau)$, which is the maximum of the fitted TF column density at the same time. The coordinate $\tilde{y}=y/R_\perp$ is the distance from the vortex axis in units of the transverse TF radius obtained from the same fit. The spatial range is limited to half the TF radius in order to highlight the print of the vortex in the column density; the effects of the condensate boundaries are almost negligible in this range.      }
\label{fig:restheo}
\end{figure}

\begin{figure}[]
\includegraphics[width=0.9\linewidth]{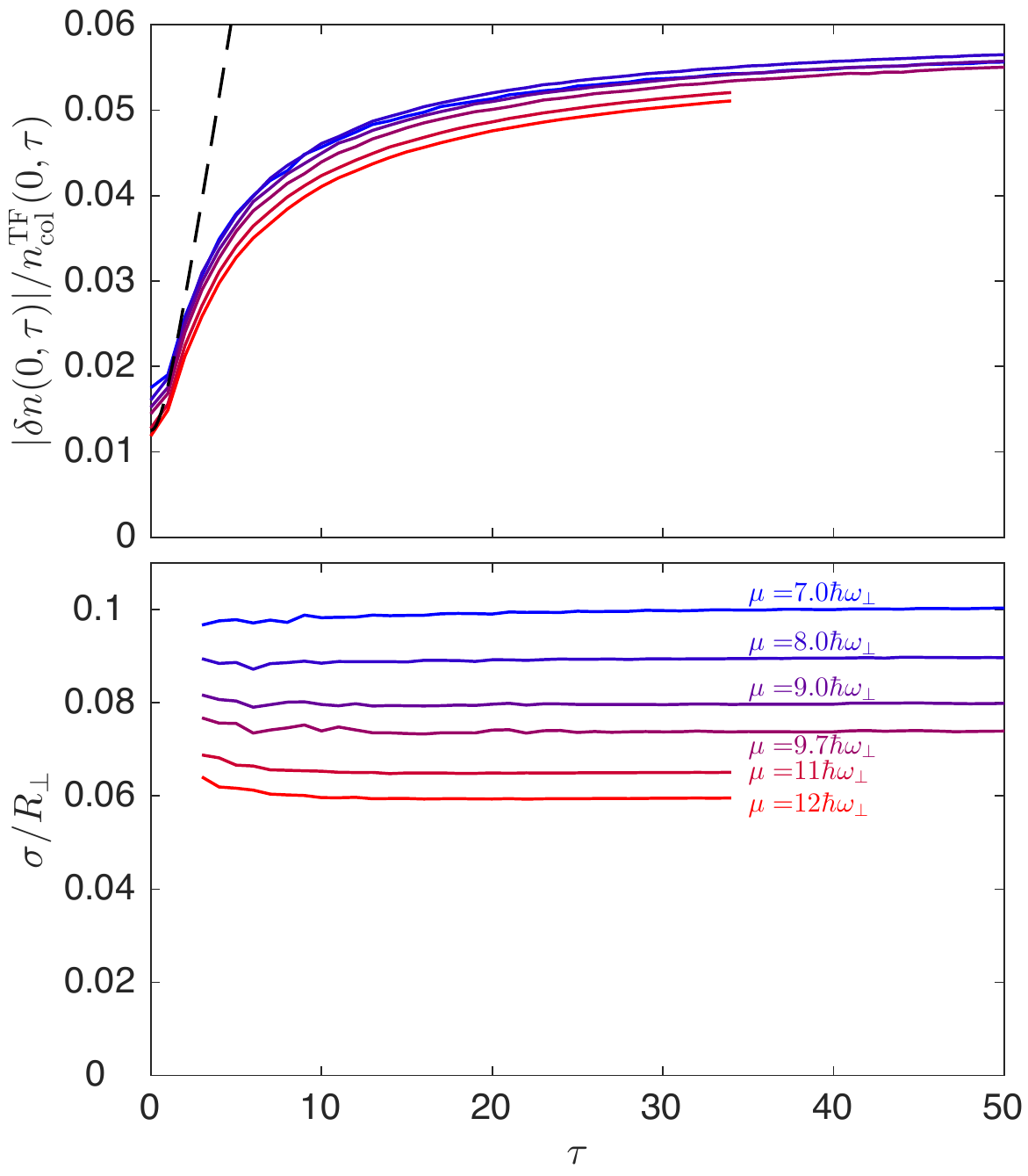}
\caption{Time evolution of the depth (top) and width (bottom) of the depletion produced by a vortex in the residual column density of expanding condensates with different chemical potentials $\mu$.
Depth and width are defined as the amplitude and the width $\sigma$ of a Gaussian fit, respectively.
As in Fig.~\ref{fig:restheo}, these parameters are normalized by the central TF column density and the transverse TF radius.
Note that to be consistent with our experiments, for the purpose of improving the fit quality, prior to fitting we average  $\delta n (\tilde{y},\tilde{z},\tau)/n^{\rm TF}_{\rm col} (0,\tilde{z},\tau)$ over different $z$ values within the interval [$-R_{\perp}/3,R_{\perp}/3$]. At very early times, $\tau\lesssim 3$, the dip in the residual is too small for the fit to quantitatively represent the vortex's characteristics. The dashed line is the prediction (\ref{eq:maxresidualvortex}) of the empty core model.}
\label{fig:depthwidth}
\end{figure}

The TF density profile (\ref{eq:rescaledTF}) is not only an accurate fitting function of the GP density distribution during the free expansion of an elongated condensate with $\mu \sim 10 \hbar \omega_{\perp}$, but the TF radii extracted from the fit also agree with the scaling solutions of (\ref{eq:b}), as well as with the analytic expressions  (\ref{eq:banalytic}), the discrepancy being less than $2$\% in all our simulations, even for long expansion times. The agreement is expected to be even better for larger values of $\mu$. This justifies the use of a TF fit to extract the residual density both in the experiments and in the GP simulations. The fit also provides the values of the TF radii and $n_0$ at any given time $t$, which can be used to rescale the coordinates and the density.  

 For comparison with experiments, the key quantity is the column density, that is, the integral of the density along the imaging axis. Let us consider a cut of the density in the $z=0$  plane and define $n_{\rm col}(\tilde{y},t) =  \int d\tilde{x} \ n(\tilde{x},\tilde{y},0,t)$, where the integral is restricted to the region where the density is positive. Using the analytic TF density, one finds 
\begin{equation}
n^{\rm TF}_{\rm col}(\tilde{y},t)=n^{\rm TF}_{\rm col} (0,t)(1 - \tilde{y}^2)^{3/2} 
\label{eq:ncolaxial}
\end{equation}
and we can finally define the residual column density as
\begin{equation}
\delta n (\tilde{y},t) =  n_{\rm col} (\tilde{y},t) - n^{\rm TF}_{\rm col}(\tilde{y},t) \,  .
\label{eq:nres}
\end{equation}
An example is shown in Fig.~\ref{fig:restheo}, where we plot $\delta n$  obtained in the GP simulation of the expansion for a condensate with $\mu= 9.7 \hbar \omega_\perp$. The figure shows that, as expected, a vortex produces a (column) density depletion whose depth is very small, i.e., only a few percent of the central column density of the condensate. It also shows that the depth increases in time during the expansion, while the width seems to remain almost constant.   In Fig.~\ref{fig:depthwidth} we show results for the depth and the width obtained in simulations of condensates with different chemical potentials,  plotted as a function of the expansion time.

These results can be qualitatively understood by using a simplified model where the GP vortex core in the initial condensate is modelled by an empty cylinder of radius $r_v = c  \xi_0$, where  $c$ is a number of order $1$ and $\xi_0$ is the healing length of a uniform condensate with density $n_0$, which is given by $\xi_0= \hbar/\sqrt{2mgn_0} = \hbar/\sqrt{2m\mu}$. The rescaled radius is $\tilde{r}_v=r_v/R_\perp= c \xi_0/R_\perp= c \hbar\omega_\perp/2\mu$. Then, let us assume that the initial expansion of the condensate is dominated by the mean-field interaction in the following sense: a segment of vortex filament near the center of the condensate expands as if it were in a uniform condensate, preserving its shape, but adiabatically following the time variation of the density of the medium around it. Hence, the vortex radius grows because the density decreases and the healing length is inversely proportional to $\sqrt{n_0}$. Meanwhile, the transverse and axial TF radii $R_\perp$ and $R_x$ grow, but with different scaling laws; such a difference is precisely the origin of the increased visibility of the vortex.  The empty-cylinder model allows us to calculate the column density, analytically taking into account all of these effects. In particular, using the scaling law (\ref{eq:banalytic}) and neglecting the $\lambda^2$ term, one can easily prove that $\tilde{r}_v$ is constant during the expansion, while the residual column density takes the form 
\begin{equation}
\delta n (\tilde{y},\tau) =  - \frac{3 \lambda \tilde{r}_v}{2}  n^{\rm TF}_{\rm col} (0;\tau)  \sqrt{1+\tau^2} 
(1 - \tilde{y}^2) \left( 1 - \frac{\tilde{y}^2}{\tilde{r}_v^2} \right)^{\frac{1}{2}}   \ ,
\end{equation}
and the normalized depth can be written as    
\begin{equation}
\frac{|\delta n (0,\tau)|}{n^{\rm TF}_{\rm col} (0,\tau)} =   \frac{3}{2} \lambda \tilde{r}_v  \sqrt{1+\tau^2}  \, .
\label{eq:maxresidualvortex}
\end{equation}
The dashed line in Fig.~\ref{fig:depthwidth} corresponds to this prediction when $c=1.6$ and $\mu= 9.7 \hbar \omega_\perp$. With the same parameters, the rescaled width of the empty cylinder is $\tilde{r}_v \sim 0.08$, which is in qualitative agreement with the data in the bottom panel of the same figure. However, the assumption of adiabaticity is expected to be valid only at short times, when the density of the expanding condensate remains sufficiently large. As the expansion proceeds, the mean-field interactions lose their strength and the velocity field gradually assumes the characteristics of a ballistic expansion \cite{Lundh98,Dalfovo00}.
The crossover from mean-field to ballistic expansion is smooth and,
for reference, we note that a spherically trapped condensate is expected to decouple at around $\tau_{\rm dec} \sim \sqrt{2\mu/\hbar\omega}$ \cite{Lundh98}, which, for $\mu = 9.7\hbar\omega$, would correspond to $\tau_{\rm dec} \sim 4$ in Fig.~\ref{fig:depthwidth}.
The full GP simulations show that the width remains approximately constant throughout the simulation, while the depth significantly deviates from the $\sqrt{1+\tau^2}$ law and saturates to a constant value deep in the ballistic regime. 

\section{Experiment \lowercase{vs.} Theory}

In this section, we compare the results of the experiments with the predictions of the GP theory for the overall shape, width and depth of the vortex in the residual column density.  

\begin{figure}
\centering
\includegraphics[width=1\linewidth]{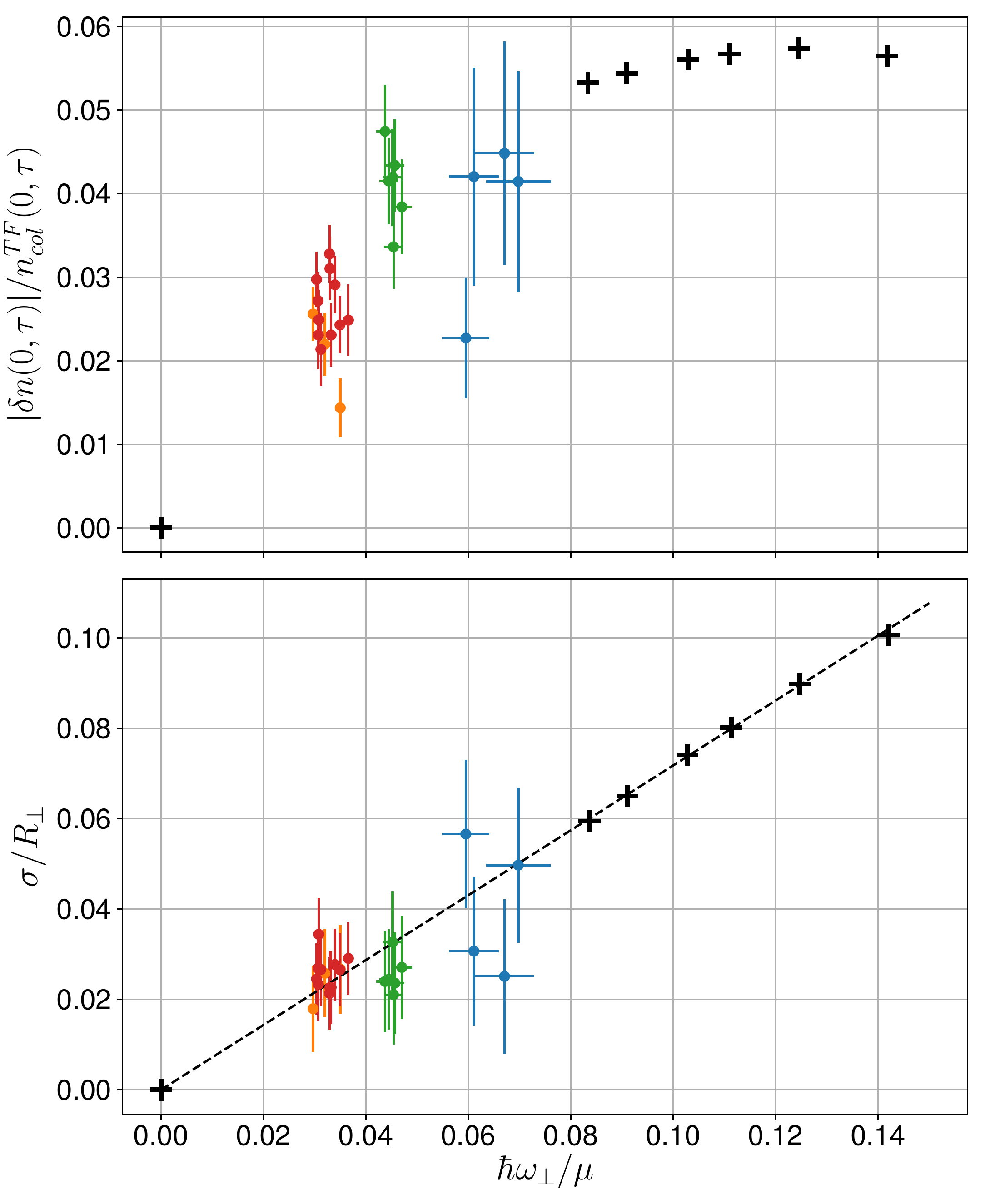}
\caption{Depth (top) and width (bottom) of the depletion produced by a vortex in the residual column density for condensates of different $\mu$. The black $+$ symbols are obtained from GP simulations for an expansion time $\tau=\omega_\perp t= 70$, corresponding to $120$~ms; the point at $1/\mu=0$ is the limit of an infinitely large condensate, where both quantities must vanish.  The dashed line in the bottom panel is the linear law $\sigma/R_\perp \sim \xi_0 /R_\perp \propto 1/\mu$ predicted by GP theory in  the TF scaling regime.  Points with error bars are the experimental data. The expansion time is $t=150$~ms (red), $t=120$~ms (green and orange) and $t=100$~ms (blue); varying $t$ in this range would change the vertical position of the experimental data by a negligible amount of the order of $1\%$.  The depth and width are calculated from Gaussian fits to both GP and experimental distributions of the residual column density by using the same procedure.   }
\label{fig:mu_scaling}
\end{figure}

The depth and the width after a given expansion time $t$ are shown in Fig.~\ref{fig:mu_scaling} as a function of $1/\mu$.  The two quantities are extracted from Gaussian fits, and normalized  by the central TF column density and the transverse TF radius as in Fig.~\ref{fig:depthwidth}. In the case of experimental data, we first select condensates exhibiting a rectilinear vortex filament near their center, at an axial distance smaller than $R_\perp/3$. We then fit the column density with the analytic TF profile, but excluding points lying within a few healing lengths of the filament. From the fit we obtain the chemical potential and the TF radii of the ``background" condensate and, by subtracting this background from the column density, we get the residual $\delta n (\tilde{y})$, where $\tilde{y}$ is taken to be orthogonal to the filament. In order to increase the signal-to-noise ratio we average the normalized depth $\delta n(\tilde{y})/n_{\rm col}^{\rm TF}(0)$ over different $z$ values within the interval [$-R_{\perp}/3,R_{\perp}/3$]. 
Moreover, if a vortex line is displaced from the center by a distance $\tilde \rho = \sqrt{\tilde x ^ 2 +\tilde y ^2 + \tilde z ^2}$, its core structure is that of a vortex in a background condensate with a density $(1 - \tilde \rho^2)$ times lower than the central density; we thus assign to the vortex a value of $\mu$ corrected by the same factor.
Finally, for long expansion times the residual external field makes the condensate slightly elliptic in the radial plane. For this reason, we use both $R_y$ and $R_z$ as independent TF radii and then we define $R_\perp=\sqrt{R_yR_z}$.  The same fitting procedure is applied to the GP density distributions, for which the condensate radius is always axially symmetric and the vortex is centered by construction. The experimental points correspond to four independent sets of data, where the cooling, evaporation, and imaging procedures are optimized for condensates with different atom numbers: red and orange points correspond to the largest condensates in our laboratory ($\mu \sim 30 \hbar\omega_\perp$, $t=150$~ms and $120$~ms), blue points are the smallest condensates in which vortices are still observable ($\mu \sim 15 \hbar\omega_\perp$, $t=100$~ms), while green points represent an old data set \cite{Bisset17} for intermediate condensates ($\mu \sim 20 \hbar\omega_\perp$, $t=120$~ms). Error bars account for statistical noise in the residual column density and for the uncertainties in the fit.   

The GP results clearly show that the rescaled width $\sigma/R_\perp$ scales linearly with $1/\mu$. This is consistent with the fact that, in the elongated geometry of our condensates, the rescaled width remains almost constant during the expansion.
Another way to understand this is to note that the in-trap width is proportional to $\xi_0 /R_\perp$, and hence to $1/\mu$, and this scaling survives after long expansion times, even deep within the ballistic regime where length ratios become frozen.
The dashed line is a linear fit to the GP points, including the limiting case of an infinite condensate at $1/\mu=0$. Figure~\ref{fig:mu_scaling} shows that the experimental data are in good agreement with the GP predictions, especially for the largest condensates, where the vortex signal-to-noise ratio is the largest.  

\begin{figure}
\centering
\includegraphics[width=1\linewidth]{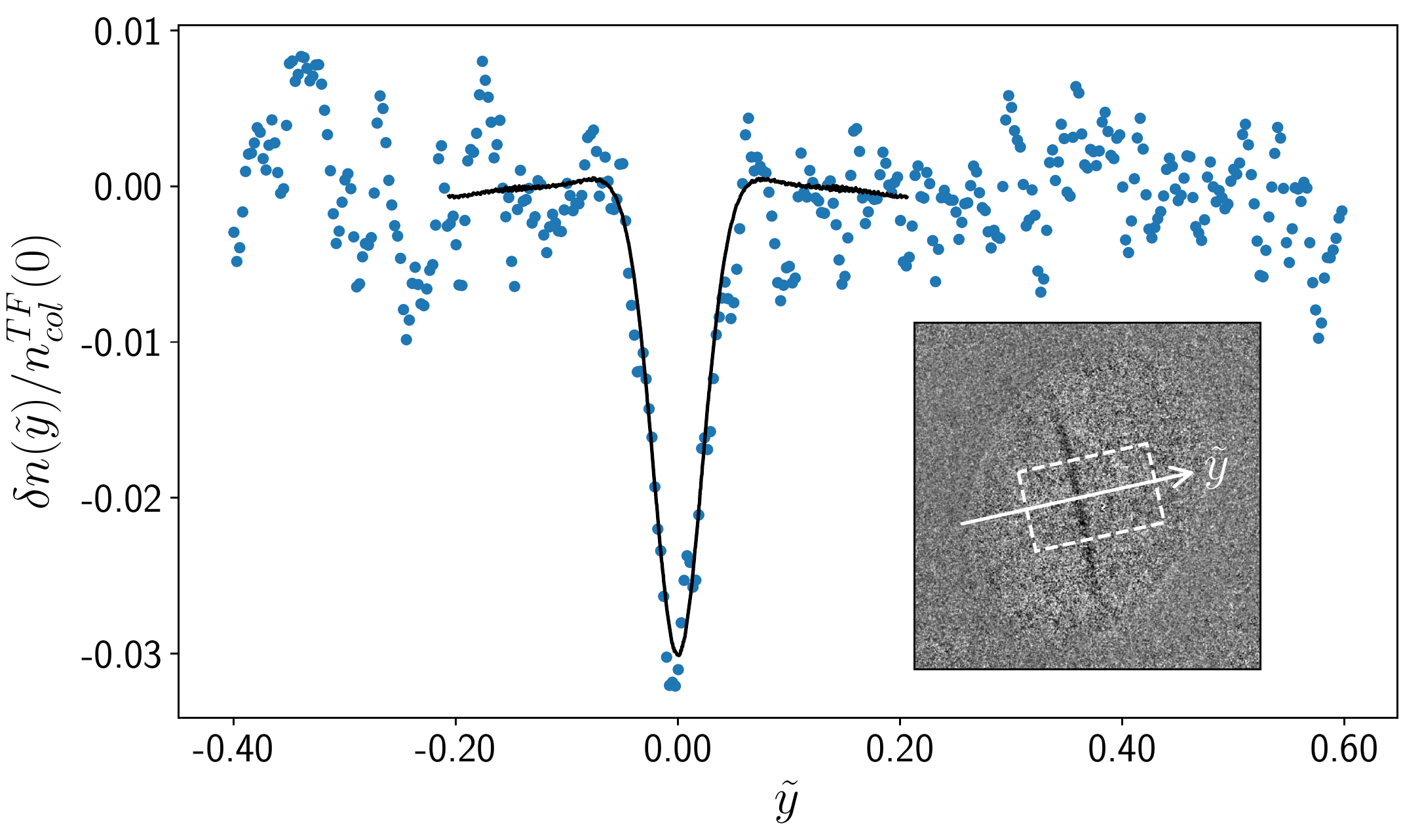}
\caption{Residual column density after $150$~ms of free expansion for a condensate with $2 \times 10^7$ atoms and $\mu=33 \hbar \omega_\perp$, containing a vortex. The inset shows the full residual column density in the $y$-$z$ plane. The quantity  $\delta n (\tilde{y},\tilde{z})/n^{\rm TF}_{\rm col} (0,\tilde{z})$ is averaged in the direction $z$ within the rectangular box and the resulting values (blue points) are plotted in the main panel as a function of the rescaled coordinate $\tilde{y}=y/R_\perp$, with $\tilde{y}=0$ at the vortex position. The solid line is the same quantity, obtained with the same fitting procedure applied to the GP residual column density of a condensate with $\mu=9.7 \hbar \omega_\perp$, after linearly rescaling its width according to  the dashed line of Fig.~\ref{fig:mu_scaling}, and reducing its depth to match the experimental value. }
\label{fig:profile}
\end{figure}

For the case of vortex depth, the GP theory does not provide any simple scaling law to compare with the experimental results considered here. The reason is that, as discussed in the previous section, the visibility of the vortex in the residual column density exhibits a nontrivial dependence on the expansion time, associated with the crossover from the mean-field dominated early stages of expansion to the later ballistic expansion dynamics. Eventually, for large $t$, the normalized depth saturates at a value weakly dependent on $\mu$ (see Fig.~\ref{fig:depthwidth}). The experimental points lie in a range fully compatible with a smooth interpolation from the GP results down to the infinite condensate limit, in the sense that any reasonable interpolating function would clearly pass through most of the experimental points, within the experimental uncertainties. 

In Fig.~\ref{fig:profile}, we show an example of vortex profile in a condensate with $2 \times 10^7$ atoms and chemical potential $\mu_{\rm expt}=33 \hbar \omega_\perp$, after an expansion time $t=150$~ms. The full residual column density $\delta n (\tilde{y},\tilde{z})$ is plotted in the inset. The quantity $\delta n (\tilde{y},\tilde{z})/n^{\rm TF}_{\rm col} (0,\tilde{z})$ is averaged in the $z$ direction within the rectangular box, and the resulting $\delta n (\tilde{y})/n^{\rm TF}_{\rm col} (0)$ is shown in the main panel of the figure as a function of $\tilde{y}$. In order to compare the experimental data with GP theory we proceed as follows. We first check that the shape of the vortex core in the residual column density of GP simulations with different values of $\mu$ is the same up to a rescaling of the width and the depth as in Fig.~\ref{fig:mu_scaling}, except for small fluctuations in the tails, which are expected to become negligible for large $\mu$. This implies that the GP profile of $\delta n (\tilde{y})/n^{\rm TF}_{\rm col} (0)$ for the experimental chemical potential $\mu_{\rm expt}=33 \hbar \omega_\perp$ should be the same as for the GP simulation for $\mu_{\rm GP}=9.7 \hbar \omega_\perp$, after rescaling the width linearly with $\mu$ (dashed line in Fig.~\ref{fig:mu_scaling}). The solid line in Fig.~\ref{fig:profile} is the resulting GP profile, where we fixed the depth to the experimental value.
There is good agreement between theory and experiment for the overall shape, including quantitative agreement for the width. The depth has good qualitative agreement if one considers that the experimental value lies within a range between the GP results for smaller $\mu$ and the trivial limit for $\mu \to \infty$, in a way that is compatible with any reasonable smooth interpolation as already shown in the top panel of Fig.~\ref{fig:mu_scaling}.
 
It is worth noticing that the optical resolution in our experiments is not limiting the comparison with theory. To check this, we convolve the GP profile with a Gaussian having a width in the range $\sigma_{\rm res} \sim 2 - 3\ \mu$m, corresponding to our optical resolution, and we find that the effects on the points in Figs.~\ref{fig:mu_scaling}  and \ref{fig:profile} are negligible (note that the vortex core in Fig.~\ref{fig:profile} has a width $\sigma \sim 30\ \mu {\rm m} \gg \sigma_{\rm res}$). The fluctuations in the experimental data, which contribute to the error bars in Fig.~\ref{fig:mu_scaling}, are dominated by photon shot-noise in the absorption images and by systematic spurious optical fringes which are not completely filtered out.   

Finally, we note that thermal atoms are not visible in our samples, which means that the temperature of the condensates is significantly smaller than the critical temperature for Bose-Einstein condensation. Nevertheless, a certain number of thermal atoms is still expected to be present in the trapped condensate, and some of them can be confined within the vortex core \cite{Coddington04}. These atoms should not be present in the vortex core after the expansion, since their kinetic energy is sufficient to separate them from the expanding condensate, leaving an empty vortex core.
In any case, our observations suggest that the effect of thermal atoms on the {\it in situ} vortex core is limited.
In fact, the good agreement that we find with GP theory (valid at zero temperature) is an indication that, if thermal atoms are present, their effects on the shape, width and depth of the vortex are negligible within the uncertainties of our experiments. 

\section{Conclusion}

In summary, we have shown that quantized vortex filaments can be observed by optical means in 3D Bose-Einstein condensates of weakly interacting ultracold atoms, at a level of accuracy which is enough to allow for a direct comparison with the predictions of the Gross-Pitaevskii theory for the width, depth, and overall shape of the vortex core. We found good agreement between theory and experiment. We have performed experiments with large condensates of sodium atoms and compared the results to those obtained in numerical simulations. In order to make the vortex visible we let the condensate expand for a long time. The expansion dynamics were included in the numerical simulations. We have shown that Thomas-Fermi scaling laws, valid for large elongated condensates, can be efficiently used to relate the observed features after expansion to the structure of the vortex core in the initially trapped condensate.   \\

\bigskip

{\bf Acknowledgments:}
We dedicate this paper to Lev P. Pitaevskii in celebration of his 85th birthday.
No words can express our gratitude for the times spent working alongside him and, of course, for his pioneering contributions to physics itself.
This work is supported by Provincia Autonoma di Trento and by QuantERA ERA-NET cofund project NAQUAS. \\

\end{document}